\documentclass[prb,amssymb,floatfix,twocolumn]{revtex4}
\usepackage{graphicx}
\begin{document}
\title{Localization of non-interacting electrons in thin layered disordered
systems}
\author{V. Z. Cerovski, R. K. Brojen Singh and M. Schreiber}
\affiliation{Institut f\"ur Physik, Technische Universit\"at Chemnitz, 09107
Chemnitz, Germany.}
\date{\today}
\begin{abstract}
Localization of electronic states in disordered thin layered systems with $b$
layers is studied within the Anderson model of localization using the
transfer-matrix method and finite-size scaling of the inverse of the smallest
Lyapunov exponent.  The results support the one-parameter scaling hypothesis
for disorder strengths $W$ studied and $b= 1,\dots, 6$.  The
obtained results for the localization length are in good agreement with both
the analytical results of the self-consistent theory of localization and the
numerical scaling studies of the two-dimensional Anderson model.  The
localization length near the band center grows exponentially with $b$ for 
fixed $W$ but no localization-delocalization transition takes place.
\end{abstract}

\maketitle

\section{Introduction}\label{1}

The one-parameter scaling theory of Abrahams et~al.~\cite{aalr} predicts that
all single-particle electronic states of non-interacting electronic systems in
two dimensions (2D) are localized for arbitrarily weak random potential has
attracted much attention in the last 25 years. This subtle effect is understood
as being due to the constructive interference between time-reversed paths which
increases the probability of returning to the original position leading to
universal conductance fluctuations~\cite{awl,ls} and corrections to
conductivity, whose divergence in $d=2$ explains the absence of
diffusion~\cite{lr,vw,gb,cs,mk,km}.  It has also been investigated numerically
for different lattice structures, e.g., square, honeycomb and triangular at the
band center ($E=0$) and away from the band center in two-dimensional (2D)
systems, but no extended states were found for the non-interacting and
zero-field case, supporting the scaling hypothesis~\cite{mo}.  There have been
more theoretical and experimental investigations to see the validity of the
conclusions of scaling theory~\cite{do,btd,udp,mcg,aks,vw}, supporting the
theory from a variety of numerical, experimental and analytical approaches in
the absence of electron-electron interactions, magnetic field and spin-orbit
interactions \cite{mk,km,lr,vw,ls,damc,aks}. 

In 3D systems on the other hand, scaling theory predicts that there exists a
second-order continuous phase transition, with a critical disorder above which
all the states are localized whereas below it extended states appear at the
band center, separated by mobility edges from the localized states in the rest
of the spectrum. This was supported by numerical~\cite{mk,km,ps,so} and
analytical investigations \cite{vw,ls}, with $d=2$ being the lower critical
dimension of the transition. The question then can be posed what happens in
thin films of finite thickness.  When such a small thickness in terms of a few
layers is introduced in an infinite 2D electron system, the degree of freedom
of the electron motion will be increased.  This thin film can be expected to
behave as a 2D system, if the Thouless length $L_{\mathrm{Th}}$, which is the
length scale up to which an electron can diffuse without inelastic collisions,
is larger than the film thickness $b$, whereas for usual thin films,
$L_\mathrm{Th}<b$~[\onlinecite{lr}].  The study of such systems should be
important because experiments are always done in thin films of finite
thickness, and some experiments \cite{lkt} which studied electron transport
with the variation of film thickness showed a fascinating role of the film
thickness in the metal-insulator transition (MIT): the authors observed an MIT
in Mo-C films at a certain thickness and found that the characteristic
parameters like localization length and dielectric constant grow exponentially
with increasing thickness near the transition.

The numerical investigations in 2D and 3D systems played an important role to
study the properties of electronic eigenstates in these systems. There have
been various numerical methods applied in these systems, e.g.~using exact
diagonalization of the secular matrices~\cite{dia}, energy-level
statistics~\cite{els}, transfer-matrix method (TMM) \cite{mk,ps,mo}, studies
of the dimensionless conductance \cite{s,smo}, etc.  The results of these
studies support the one-parameter scaling hypothesis.

Recently~\cite{bd}, there was a discrepancy between numerical results obtained
using the TMM and an analytical analysis of thin films of finite thickness
based on the self-consistent theory~\cite{vw,lr}.  Numerically, an MIT induced
by film thickness was found whereas the analytical results showed no metallic
phase if the thickness of the film is finite~\cite{bd}.  The discrepancy is
rather surprising since previous studies of 2D systems~\cite{mk,km,mo,zbk} showed
good agreement between the analytical and numerical results.

With this motivation we address numerically the problem of localization in thin
disordered systems of noninteracting electrons using the standard TMM and
finite size scaling (FSS)~\cite{mk,ps} for non-interacting electrons in zero
magnetic field as described by the Anderson model of localization (defined
below), in systems of considerably larger sizes than in~Ref.~[\onlinecite{bd}]
and compare it with the analytical results.  To this end we first note that the
analytical expression for the localization length $\xi$ from~[\onlinecite{bd}]
can be expressed as
\begin{equation}\label{eq:bro}
 \xi = F\sqrt{b'\over\sinh b'}\exp(b'\,C/W^2),
  \end{equation} 
where  $C, F$ are non-universal constants, $W$ is the disorder strength and
$b'$ is the film thickness.  The expression is valid for $\lambda_F\ll\ell\ll L
<L_{\mathrm{Th}}$ and  $\lambda_F\ll b'\ll L$, where $\lambda_F$ is the Fermi
wavelength, $\ell$ the elastic mean-free path, and $L$ is the system size.  In
infinitely thick films ($b'\to\infty$), on the other hand, a crossover from
2D to 3D takes place where one can get both a metallic and an insulating phase
depending on the disorder strength.  

In section II, we describe the model and the notations; the TMM and FSS are
discussed in section III;  obtained numerical results are presented in section
IV and some conclusions based on our numerical results drawn in section V.

\section{The Model}

The Hamiltonian we have used for our numerical calculations is the Anderson
model of localization given by
\begin{equation}\label{eq:Anderson}
  H=\sum_{r}^N\epsilon_r|r\rangle\langle r|-
    t\sum_{\langle r,r'\rangle}^N |r\rangle\langle r'|,
\label{H}
\end{equation}
where $|r\rangle$ is a tight-binding state at site $r$. The sites $r$ form a
simple cubic lattice with  $N=M\times b\times L$ atoms where $L$ is the length,
$M$ is the width, and $b$ is the thickness of the lattice, here measured in the
number of layers of the system, in units of the distance $a$ between the
layers.  Since the only relevant microscopic scale in the continuum model that
gives Eq.~(\ref{eq:bro}) is $\ell$, we compare the
length scales of the continuum and Anderson model {\it via} the relation
$b'=\alpha b$, where $\alpha\equiv a/\ell$ is a non-universal number which
needs to be taken into account when comparing analytical and numerical results.

We use open boundary conditions along the $b$ (thickness) direction and
periodic boundary conditions along the $M$ (width) direction. The on-site
potentials $\epsilon_r$ are taken to be randomly distributed in the interval
$[-W/2,W/2]$;  $\langle r,r'\rangle$ denotes that the hopping $t$ is restricted
to the nearest neighbors and the energy units are chosen by setting $t=1$.

We assume that this model describes the thin layered system if $b\ll M$.  When
$b$ becomes comparable to $M$ at small $W$, one should expect a crossover from
insulating towards metallic behavior after FSS, and it is therefore important
to have $b/M$ as small as possible in order to stay in the limit of finite
thickness. We therefore attribute the appearance of the metallic-like behavior
in the numerical simulations of Ref.~[\onlinecite{bd}] to the large values of
$b/M$ used.  Indeed, for $b=6$, for instance, as $M$ was changed from $2$ to
$15$, the cross-section of the bars changed from rectangular to square to
rectangular shape, and the scaling analysis for such geometries should not be
compared with the analytical results of~[\onlinecite{bd}].  Instead, we focus
on systems (\ref{H}) with $M\gg b$.  In the case of the thickest system studied
here, for example, the scaling analysis is done for $b=6$ and $M=23,33,47,65$
as we discuss in detail below.

\section{One-parameter finite size scaling}

In the TMM one determines localization properties by a scaling analysis of the
Lyapunov exponents (LE) of the transfer matrices of quasi-1D bars. 

The largest length scale $\lambda_{Mb}$ is obtained from the smallest LE,
$\lambda_{Mb}\equiv 1/\gamma_{Mb}$. The accuracy of these $\lambda_{Mb}$ is
determined from the variance of the changes of the exponents in the course of
the iteration, and the relative error of $\lambda$ is 1\% in our calculations.
For $b=1$ our system is a 2D strip of width $M$ and length $L\gg M$, while for
$b=M$, it becomes a 3D bar with square cross-section of $M\times M$ atoms and
length $L$.

We check the one-parameter scaling theory by the FSS analysis~\cite{mk} of
the renormalized decay length, $\Lambda_{Mb}=\lambda_{Mb}/M$, for each $b$
individually by expressing it as
\begin{equation}
  \Lambda_{b}(M,W)=f_b\left(\xi_b(W)/M\right),
\end{equation}
where $\xi_b(W)$ is the scaling parameter which can be identified as the
localization length in the insulating regime and the resistivity in the
metallic regime of the infinite system~\cite{mk} and $f_b(x)$ should be, if the
one-parameter hypothesis holds, a universal function $f(x)$ independent of $b$.
The scale of $\xi$ is determined from the condition that $f_b(x) = x$ when
$x\to 0$~[\onlinecite{mk}], which corresponds to expressing $\xi$ in units of
$a$.  The numerical determination of $f_b(x)$ is done by the method based on
Ref.~[\onlinecite{mk}], and the thus obtained scaling functions for different
$b$ are compared to check for the universality.  If there is an MIT, then
$\Lambda_{b}$ for given $b$ will have two branches, one which goes to zero as
$M$ increases (indicating localization of electronic eigenstates) and the other
which diverges as $M$ increases (indicating extended eigenstates).  

\begin{figure*}
\includegraphics*[width=6.21in]{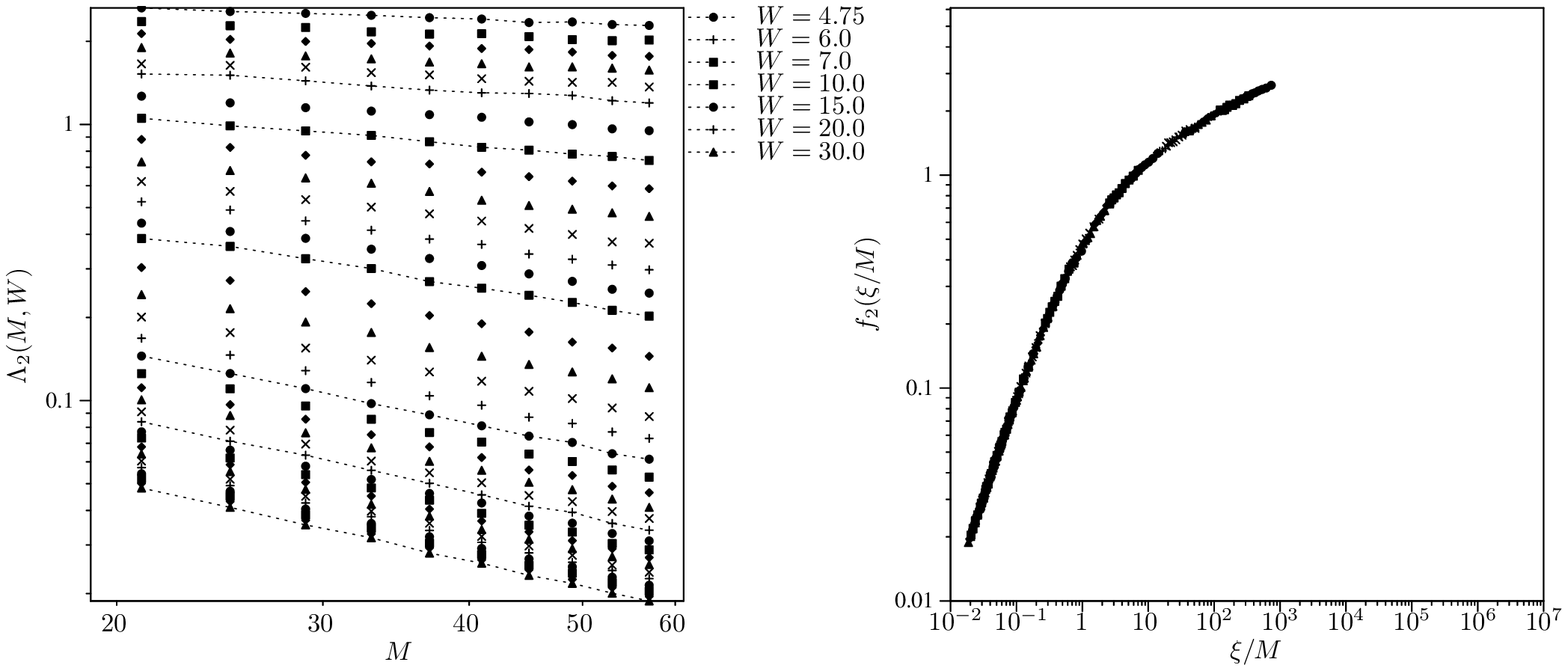}
\includegraphics*[width=6.21in]{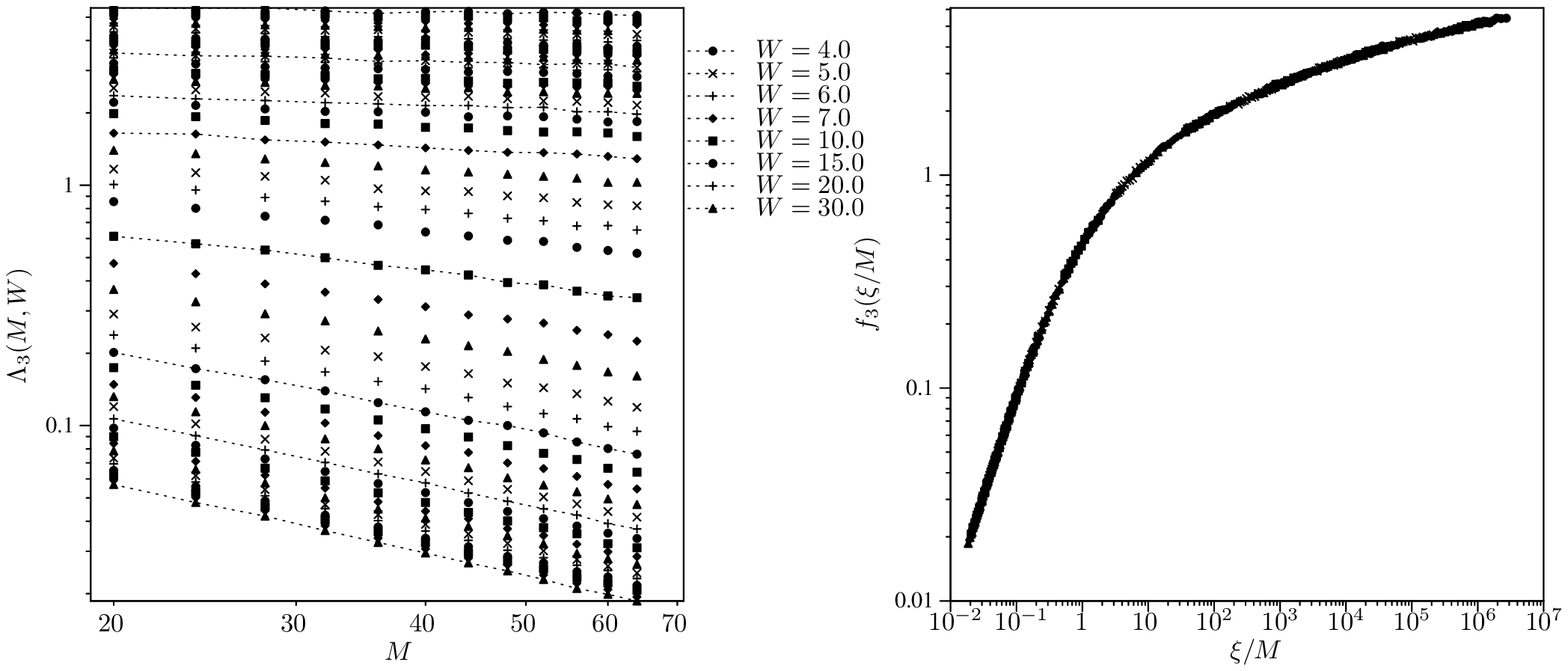}
\includegraphics*[width=6.21in]{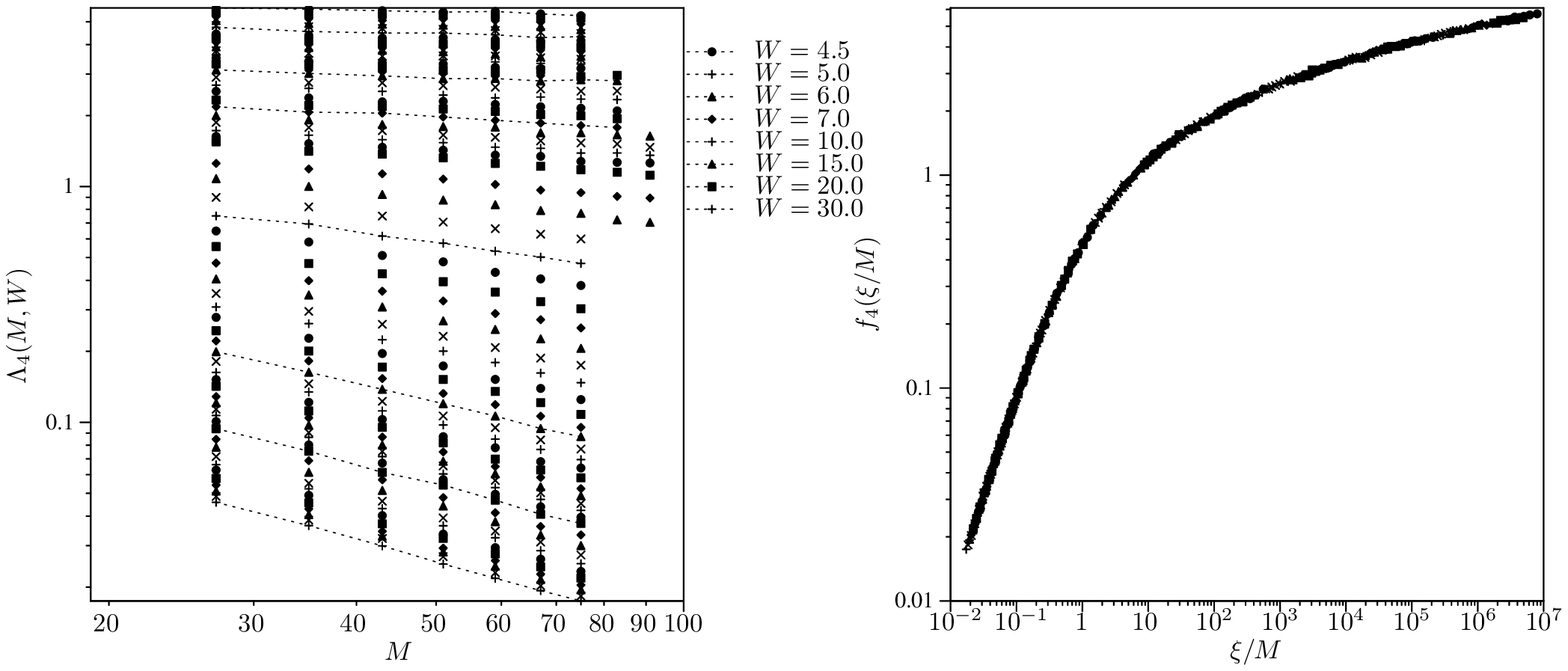}
\caption
{Plot of the renormalized decay length $\Lambda_{b}(M,W)$ (left panels),
scaling function $f_b$ (right panels) for systems of width $M$ and $b=2,3,4$
layers, from top to bottom. Dotted lines connect points for $W$ values
indicated in the corresponding legend.
}
\label{fig:all}
\end{figure*}

\begin{figure*}
\includegraphics*[width=6.21in]{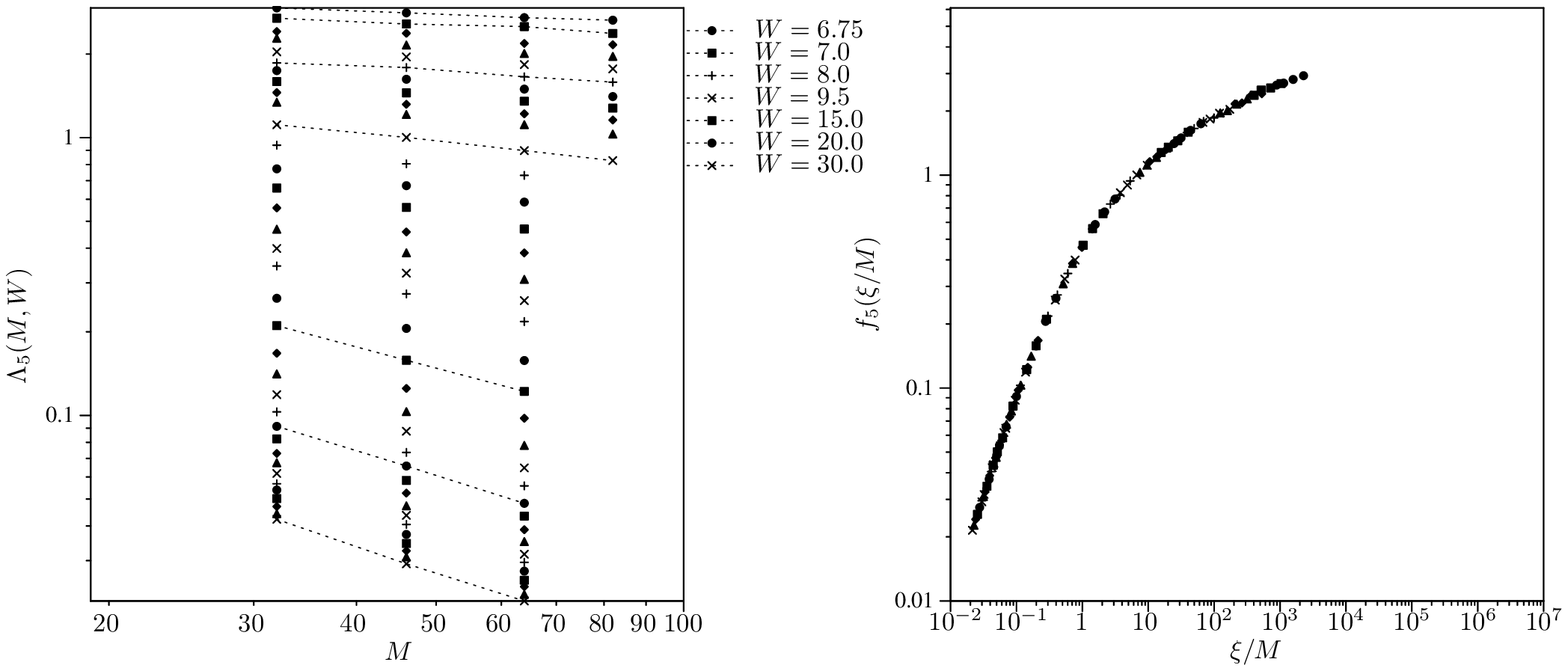}
\includegraphics*[width=6.21in]{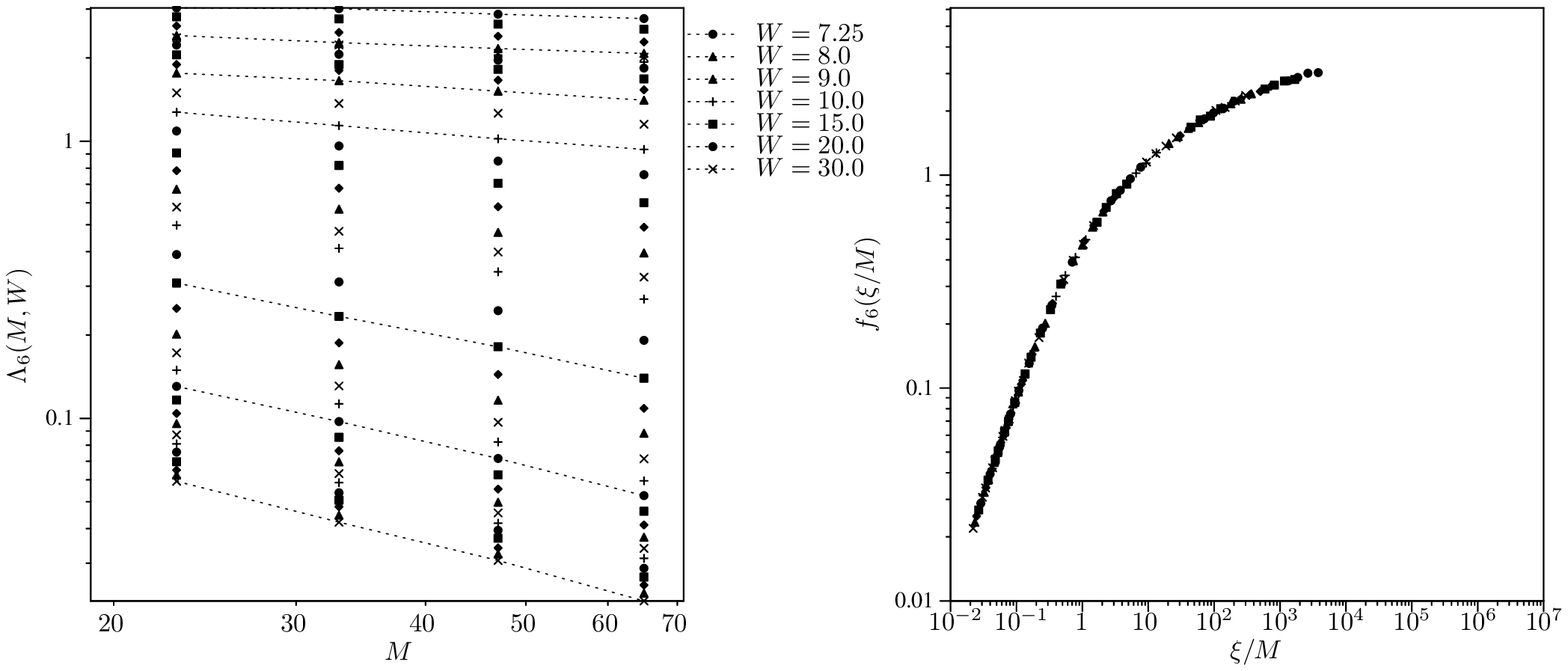}
\caption{Same as Fig.~\ref{fig:all}, for $b=5$ and $6$ layers, respectively.}
\label{fig:rem}
\end{figure*}

Fig.~\ref{fig:all} shows $\Lambda_{b}(M,W)$ as a function of strip width $M$
for various $W$, the scaling function $f_{b}(\xi_b(W)/M)$, as well as the
localization length $\xi(W)$ for systems with $b=2,3,4$ layers, and
Fig.~\ref{fig:rem} shows the same quantities for systems with $b=5,6$ layers.
Our results for $b=1$ are not shown since they are in agreement with results
obtained before~\cite{zbk}.  For larger $W$ values, $\Lambda_{Mb}$ decreases as
the width of the system $M$ increases indicating localization, while for
smaller $W$ values the decrease becomes much weaker.  

In addition to the results presented in Figs.~\ref{fig:all} and \ref{fig:rem},
we have calculated $\Lambda_b$ for $W$ values as small as $2,2,4,4,4$ for
$b=2,3,4,5,6$, respectively, but no metallic behavior was found.  Instead,
$\Lambda_b(M)$ become approximately constant for smaller $W$ values, which is
in qualitative agreement with 2D being the lower critical dimension of the
localization-delocalization transition.

\begin{figure}
\includegraphics*[width=3.0in]{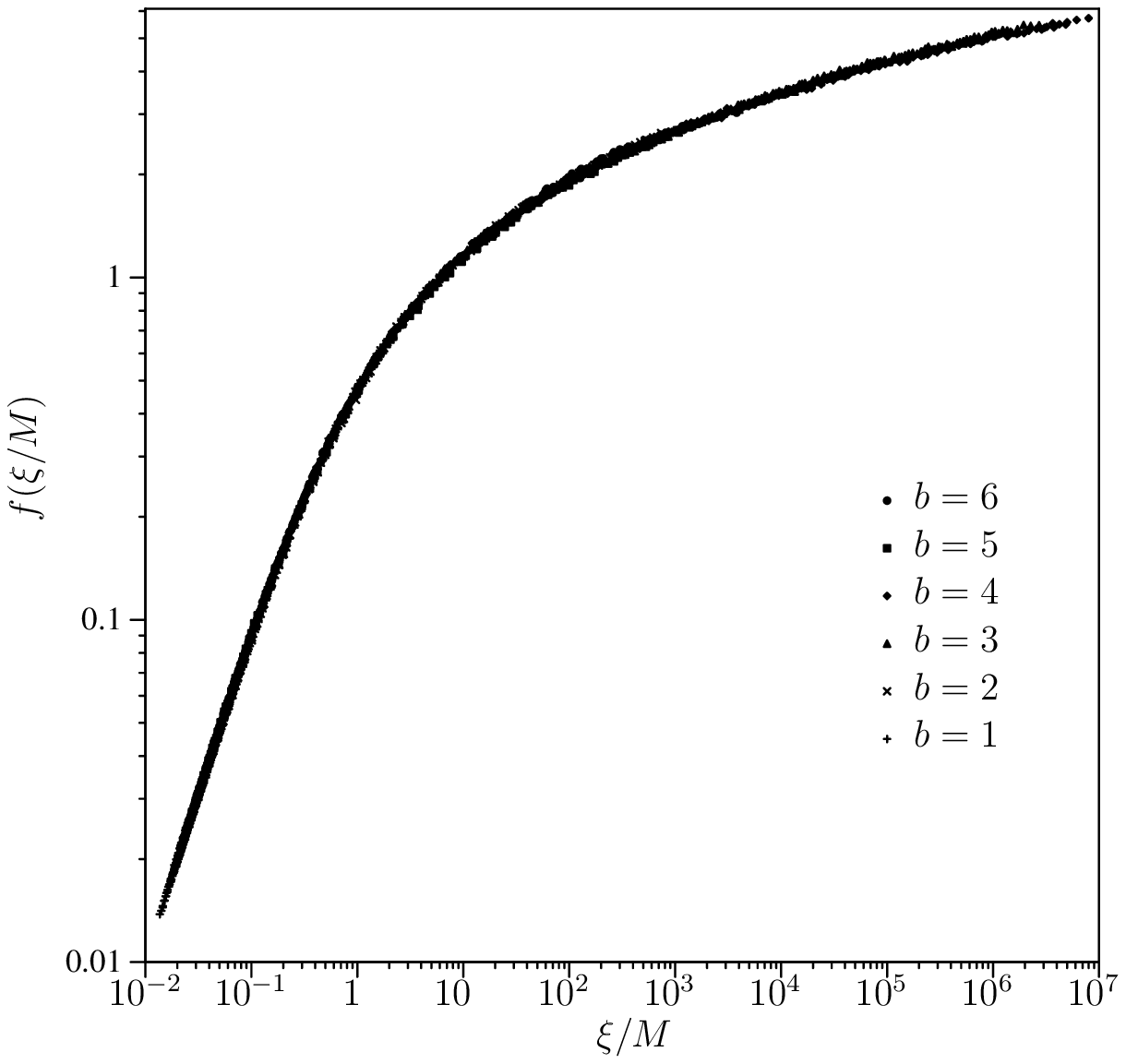}
\caption{Plot of the scaling functions $f_b(\xi/M)$ for $b=1,2,\dots,6$ from
Figs.~\ref{fig:all} and \ref{fig:rem}.}
\label{fig:f}
\end{figure}

To see the dependence of the scaling function $f_b$ on $b$, we first plot all
the obtained results for $f_b$ in Figs.~\ref{fig:all} and \ref{fig:rem} using
the same range on the axes for easier comparison, and then show all of the
obtained $f_b$ together in Fig.~\ref{fig:f}, without any rescaling or shifting
of individual curves, including the curve for $b=1$.  The figure shows that
$f_b$ does not depend on $b$, therefore supporting the universality of the
scaling function $f(x)$.  Since it has only one branch corresponding to the
localized states it is also in agreement with the one-parameter scaling
hypothesis for disorder strengths $W\ge 4.75,4.0,4.5,6.75,7.25$ for
$b=2,3,4,5,6$, respectively.  For $b=1$, one parameter-scaling was
found~\cite{mk,zbk} to hold for $W>2$, and the previously obtained~\cite{mk}
scaling function $f(x)$ for 2D system agrees within error bars with the $f(x)$
obtained here for $b\ge 1$, furthermore supporting the universality of the
result.

\begin{figure*}
\includegraphics*[width=6.2in]{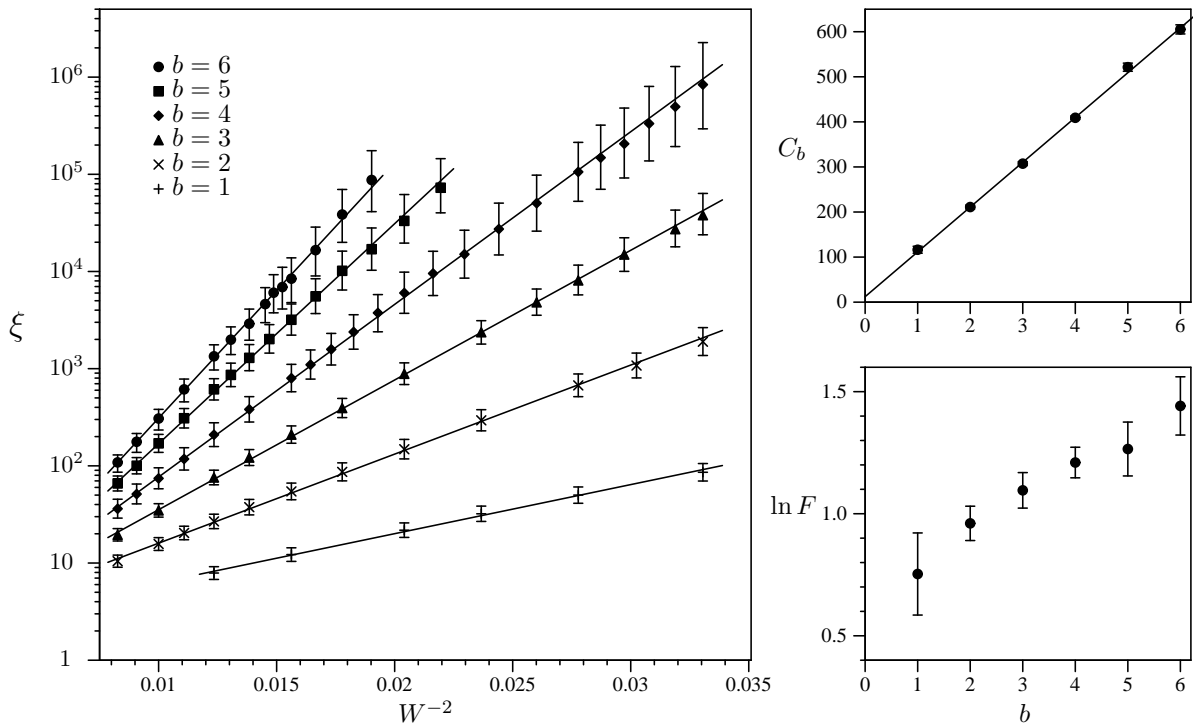}
\caption{The left panel: Dependence of the localization length $\xi_b$ on
$W^{-2}$ .  Straight lines are fits to the expression $\xi_b(W) = A_b
\exp(C_b/W^{2})$.  The rightmost point for each $b$ value correspond to the $W$
of the topmost curve of the corresponding panel in Fig.~\ref{fig:all} and
\ref{fig:rem}.  The right panels: dependence of the fitting parameters on the
thickness $b$.}\label{fig:xi}
\label{func-renorm}
\end{figure*}

To compare the dependence of the localization lengths on $W$ to the analytical
result~(\ref{eq:bro}), we fit the obtained $\xi_b(W)$ to the form
\begin{equation}\label{eq:fit}
 \xi_b(W) = A_b \exp\left({C_b\over W^{2}}\right),
\end{equation}
as shown in the left panel of Fig.~\ref{fig:xi}.  The upper and lower right
panels of the same figure show, respectively, the dependence of $C_b$ and
$F_b\equiv A_b\sqrt{\sinh b/b}$ on the number of layers $b$.  The upper right
panel of the figure also contains the mean least-squares linear fit of $C_b$,
which gives $C_b = C\,b + C_0, C=99\pm 4, C_0 = 10\pm 15$.  The obtained value
$C_0\approx 0$ is an additional check that the numerical results are in
agreement with Eq.~(\ref{eq:bro}) as far as the $W^{-2}$ dependence in the
exponent is concerned.

\begin{figure}
\includegraphics*[width=2.6in]{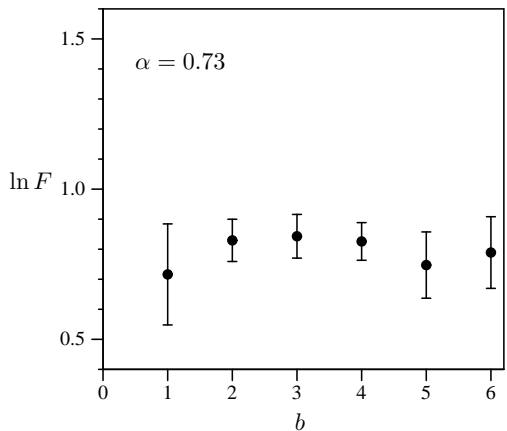}
\caption{Dependence of $\ln F$ on $b$ for $\alpha=0.73$.
}\label{fig:lnF-renorm}
\label{func}
\end{figure}

The obtained values of $\ln F$ depend on $b$, which seems to be in
contradiction with Eq.~(\ref{eq:bro}) but can be explained by taking into the
account the scale factor $\alpha$, and Fig.~\ref{fig:lnF-renorm} shows $\ln F$,
where $F\equiv A_b\sqrt{\sinh\alpha b/\alpha b}$ for $\alpha=0.73$, when $\ln
F$ becomes approximately constant, and $\alpha=0.73\pm 0.13$, where the error
bar is estimated from the range of $\alpha$ values for which the values of $\ln
F$ for different $b$ remain within error bars among themselves. This result can
be interpreted as that the effective $\ell$ of the lattice
model~(\ref{eq:Anderson}) in the range of $W$ studied is of the order of
$1.4\pm 0.2$ lattice constants.

As an additional check of the validity of the above results we notice that the
obtained value of $C$ is in agreement with the value $C\approx 100$ obtained in
the numerical studies of the 2D Anderson model~\cite{mk,zbk}, i.e.~for $b=1$,
even though the method by which we determine $C$ relies essentially on the
study of systems with $b>1$.

\section{Conclusions}
Thin quasi-1D bars with rectangular cross section of $M\times b$ and length
$L\gg M$ atoms were studied using the TMM and FSS of the inverse of the
smallest Lyapunov exponent, for $b=1,2,3,4,5,6$ and $M\gg b$.  The obtained
localization lengths were found to be in a good agreement with the analytical
results.  Despite the exponential increase of the localization length as the
thickness increases, we find no signs of extended states for the disorder
strengths studied, in agreement with the one-parameter scaling hypothesis.

\section*{Acknowledgement}

One of us (B.K.R.S.) would like to thank to Prof.~Deepak Kumar for valuable
discussions.

\end{document}